\documentclass{sf2a-conf2015}
\usepackage{graphicx}
\usepackage{hyperref}
\usepackage[]{natbib}
\usepackage{epstopdf}

\def\BibTeX{{\rm B\kern-.05em{\sc i\kern-.025em b}\kern-.08em
    T\kern-.1667em\lower.7ex\hbox{E}\kern-.125emX}}

\begin{document}

\TitreGlobal{SF2A 2015}


\title{Photometric analysis of the corona during the 20 March 2015 total solar eclipse: density structures, hydrostatic temperatures and magnetic field inference.}

\runningtitle{Photometric analysis of the solar corona and helium shells}

\author{C. Bazin}\address{Institut d'Astrophysique de Paris - UPMC and Sorbonne Universite, 75014 Paris}

\author{J. Vilinga $^{1,}$}\address{University of Luanda, Angola}

\author{R. Wittich}

\author{S. Koutchmy $^1$}

\author{J. Mouette $^1$}

\author{C. Nitschelm}\address{Unidad de Astronom\'ia, Facultad de Ciencias B\'asicas, Universidad de Antofagasta, Antofagasta, Chile}

\setcounter{page}{237}


\maketitle


\begin{abstract}
We present some new accurate CCD photometry analysis of the white light solar corona at the time of the last 20 March 2015 total eclipse (airborne observations on a Falcon 7X and at ground-based Svalbard). We measured coronal brightness profiles taken along radial directions from 1.001 to 3 solar radii in the northern, southern and equatorial regions, after removing the F corona and the sky background. These studies allow to evaluate the density gradients, structures and temperature heterogeneity, by considering the Thomson scattering in white light of the K corona and also emissions of the EUV Fe XII 193A (1 to 2 MK) and Fe XI 171/174 (lower temperature) simultaneously observed by SDO/AIA and SWAP Proba2 space missions. Some dispersion between the regions is noticed. The limitation of the hydrostatic equilibrium assumption in the solar atmosphere is discussed as well as the contribution of the magnetic field pressure gradients as illustrated by a comparison with the model stationary magnetic corona from Predictive Sc. Inc. These results are compared with the results of the quieter 2010 total solar eclipse corona analyzed with the same method. This photometric analysis of the inner and intermediate white light corona will contribute to the preparation of the Aspiics/Proba 3 flying formation future coronagraphic mission of ESA for new investigation at time of artificial eclipses produced in Space. Note that Aspiics will also observe in the He I D3 line at 5876 A, and will record intensities of the Fe XIV line 5303A simultaneously with the analysis of the orange white- light continuum, including precise polarimetry analysis.
.
\end{abstract}



\section{Introduction}
We intend to compare some heterogeneity typical values of the temperature and of the density in the corona, using white light observations from recent solar total eclipses, on 22 July 2010 (before maximum of activity), and at the recent 20 March 2015 total eclipse taken well after the maximum of activity. The choice of a five years interval corresponds to an extended period of the activity cycle of the corona. We performed intensity profiles to deduce the brightness in units of the mean solar disc intensity, along the North and the South poles and along equatorial regions, from the limb to 3 solar radii. The simultaneously observed bright star XZ Pisces was identified in images of 2015 and was used as a photometric reference. Its magnitude is 5.6 and sampled on 80 adu in green. The background and sky was subtracted. The exponential decays were applied on the brightness’s profiles to deduce the scale heights assuming an hydrostatic equilibrium. Also we performed Baumbach fitting to compare some values of the power law at distances beyond 2 solar radii. We found that the following polynomial law  $P1*\rho^{-17}+P2*\rho^{-7}+P3*\rho^{-3}$ was the best suitable fit on the brightness’s profiles.

\section{Brightness profiles and flash spectra analysis}

The adjustment with the Baumbach polynomial equation gives a better fitting than with the exponential decay, see also November and Koutchmy, 1996. However, after a further processing (F corona and sky background substraction), we used the exponential decay to compare the scale height in 2010 and 2015 in polar and equatorial regions in order to straightforwardly deduce the hydrostatic temperatures, see Figure 1.
These results show a different behavior of the coronal structure for 2015 compared to 2010. The averaging of the North South Polar Regions and the East and West equators allows to better estimate the gradients. The arrays in figure 2 summarize the results after using an exponential decay for the scale heights and hydrostatic temperatures. The coefficient values with the Baumbach polynomial fitting are indicated for evaluating the density. Comparing the helium shells extensions obtained from the flash spectra and space borne EUV images made at the same time allows to analyse cold and hot plasma and the activity in the inner parts of the corona, see Fig. 3.
Finally, the following figure 4 aims at showing that the low First Ionization Potential FIP line of Ti II and He I and He II shells seen in emission and the enhanced brightening indicates that these elements could be  ionized in lower altitudes above the photosphere, and close to the Temperature minimum regions. This is done by using high cadence CCD 12 bit imaging, and in 2015 two channels were used: one at 15 and the other at 214 flash spectra per second, with two computers for the simultaneous acquisitions and GPS.


\begin{figure}[ht!]
 \centering
 \includegraphics[width=1.0\textwidth,clip]{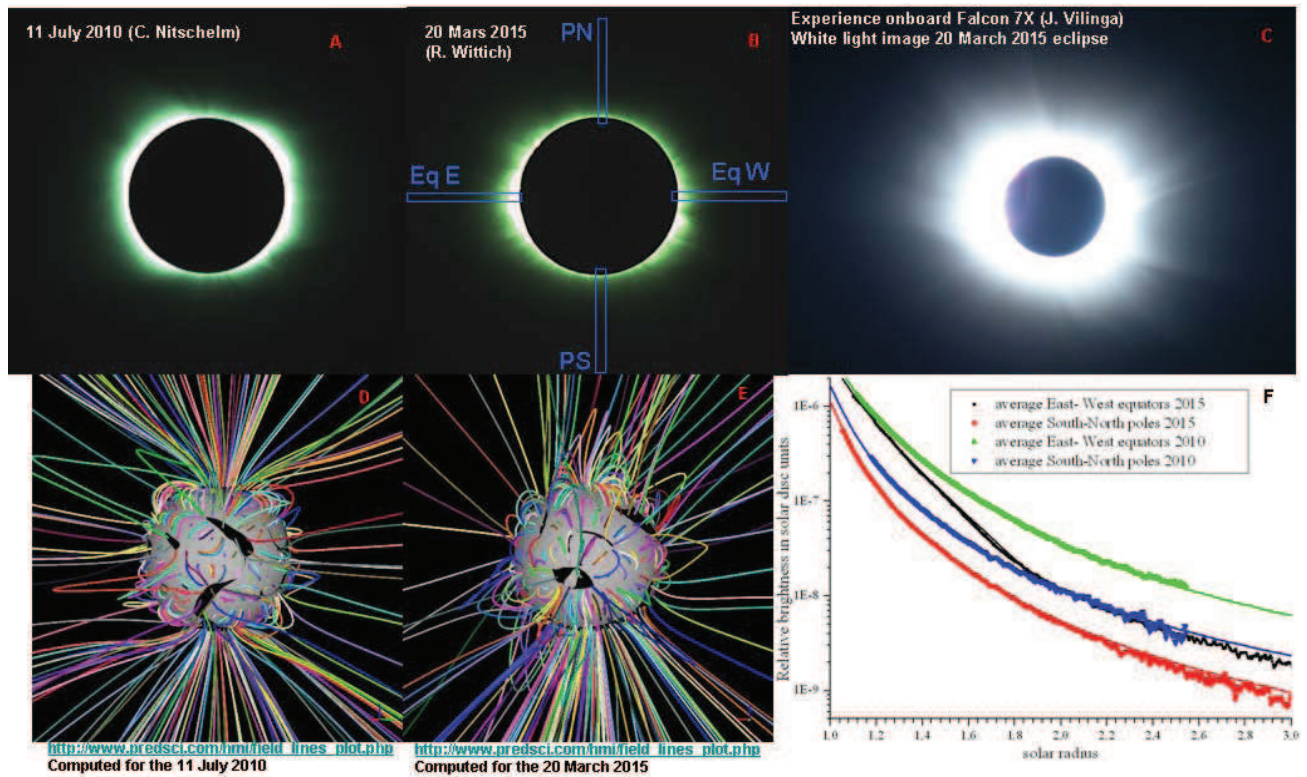} 
  \caption{From left to right and top to bottom are shown: (a) is the white light images of the total solar eclipses in 2010 (24 ms of exposure time) and (b) is the corona in 2015 (30 ms of exposure time) to perform the radial brightness profiles. (d) and (e) are the set computed magnetic field lines from the date base of \url{http://www.predsci.com/hmi/field_lines_plot.php} and (f) are the brightness profiles, near equators and poles for both 2010 and 2015}
  \label{author1:fig1}
\end{figure}
\begin{figure}[ht!]
 \centering
 \includegraphics[width=1.0\textwidth,clip]{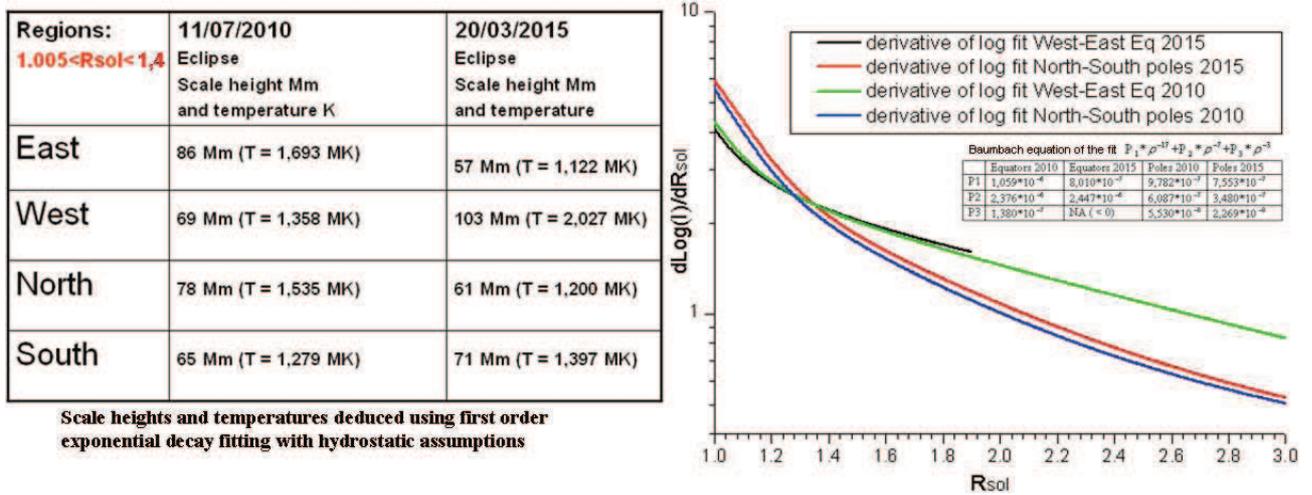}
  \caption{Scale heights and associated temperatures deduced from the derivative of the logarithm of the Baumbach fit of the brightness radial profiles in white light.
The values in the array indicates the Baumbach coefficients of the polynomial after fitting.
}
  \label{author1:fig2}
\end{figure}

\begin{figure}[ht!]
 \centering
 \includegraphics[width=0.8\textwidth,clip]{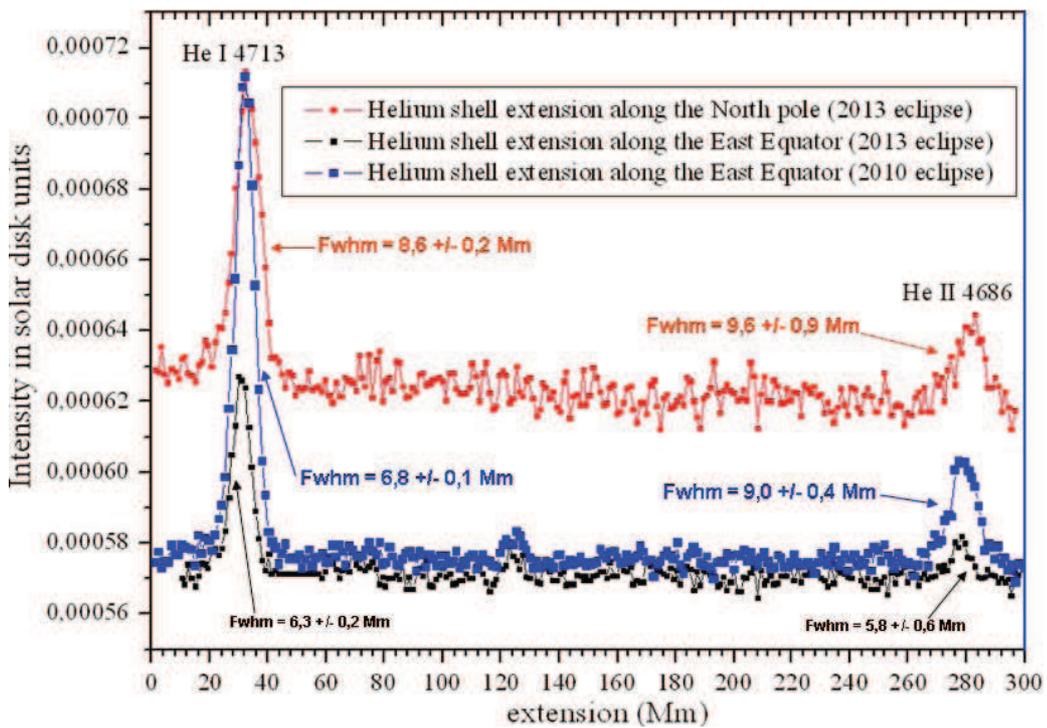}
  \caption{Helium shell extensions deduced from 2010 and 2013 3D flash spectra in optically thin layers (cool lines) but sensitive to the hot ambient corona and especially He II which is a high FIP 54 eV. In 2015 the clouds masked the end of C2 and the beginning of the totality. The scale is 0.12 Angstrom/pixel, and 1.15 Mm/pixel}
  \label{author1:fig3}
\end{figure}

\begin{figure}[ht!]
 \centering
 \includegraphics[width=1.0\textwidth,clip]{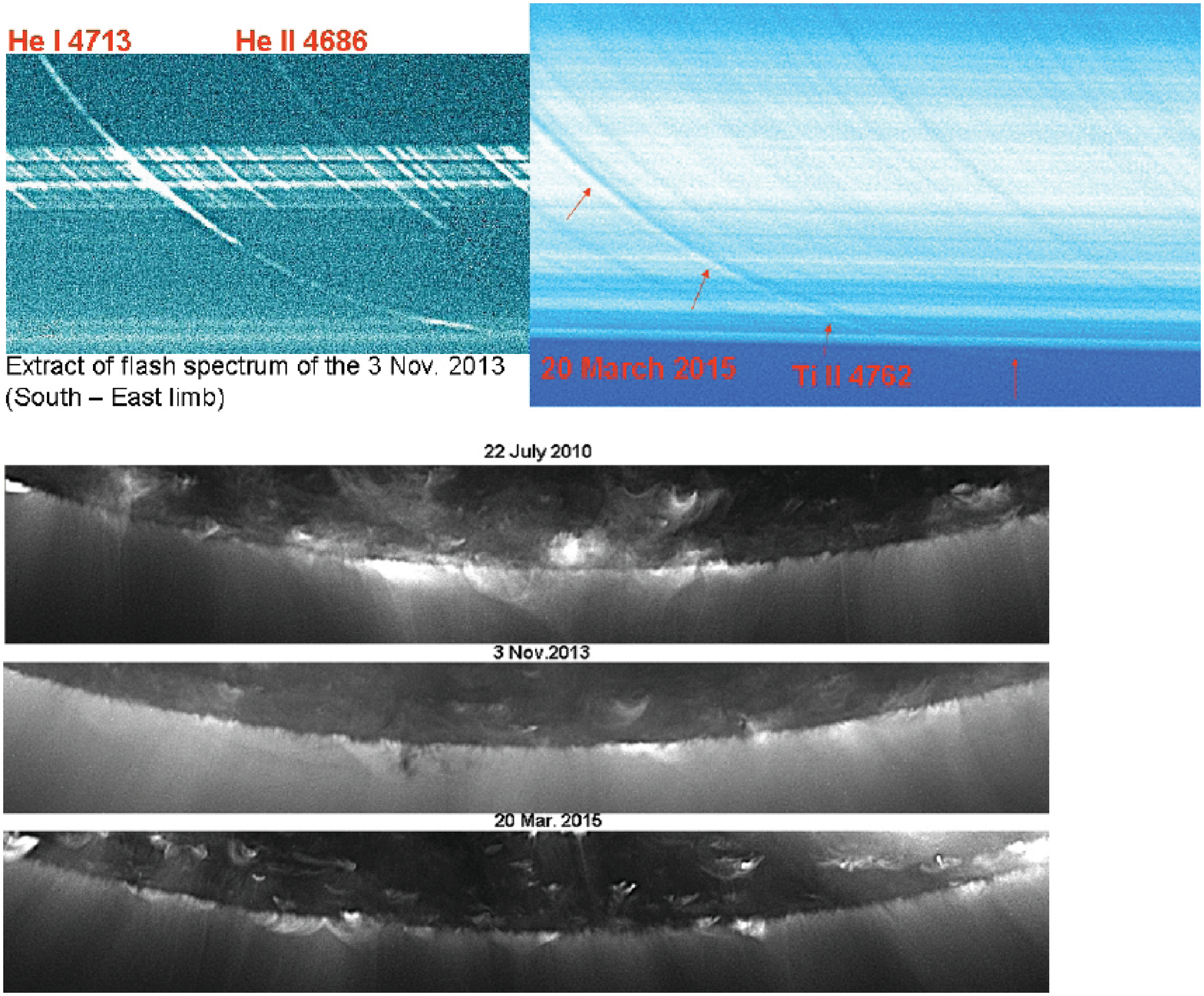}
  \caption{Extracts of flash spectra sequences of 2013 and 2015 solar eclipses for analysing the He I, He II and Ti II lines formation close to the continuum defining the solar edge, in the region of the temperature minimum. The Ti II line 4762 is a low FIP 6.83 eV. Bottom are shown some coronal holes images in the 2MK Fe XII 193 line of SDO/AIA seen at the time of the 2010, 2013 and 2015 solar eclipses, for comparing the magnetic structures, spicules and jet seen in dark (cold plasma) on the limb
}
  \label{author1:fig4}
\end{figure}

\vspace{-0.75cm}

\section{Discussion and Conclusions}

The adjustments with the Baumbach polynomial equation fit better than with the exponential decay (see figures 1 and 2). Nevertheless, we used the exponential decay to compare the scale height in 2010 and 2015 in polar and equatorial regions.
These results show a different behaviour of the coronal structure from 2010 and 2015. The averaging of the North South polar regions and the East and West equators allows to better estimate the gradients. The changing of the slope near 1.2 solar radii in 2010 and 1.4 solar radii in 2015 indicates some temperature difference and this could indicate that hot plasma penetrate at lower altitudes with a more active Sun but this will be studied in more details
The low FIP elements of Ti II lines and the high FIP He I and He II shells, seem to be ionised in the same low altitudes regions above the photosphere (T min) thanks total eclipses conditions. We compare also with coronal holes seen in 2010, 2013 and 2015, for evaluating the scales, extensions of magnetic structures, spicules and jets seen in dark on the limb (see Tavabi 2015).
We found some small differences and tendency between the 2010 and the 2015 corona. The open lines of force around the Sun show more deviation in equatorial regions in 2015 than it was in 2010. The tendency of the temperature gradients seems higher in 2015 in equatorial regions than in 2010, but the deviations are low. the extension of the helium shells also show a small increase in 2013, than it was measured in 2013. In conclusion, we found that these variations seem to be correlated, and the study of the low altitudes plays an importance using photometry, high cadence CCD flash spectra (using slit and slitless spectrographs), for studying the ionisation processes in association with magnetic activity. More analysis will be performed.


\bibliographystyle{aa}  
\bibliography{sf2a-template} 

\begin{thebibliography}{}

\bibitem[Baumbach(1937)]{Baumbach1937}
Baumbach 1937, S. Astron. Nachrichten, 263, 121

\bibitem[Bazin(2013)]{Bazin2013} 
Bazin, C., Koutchmy, S. and Tavabi, E. Journal of Advanced Research, Issue 3, 2013, 307

\bibitem[Lebecq(1985)]{Lebecq1985}
Lebecq, C., Koutchmy, S. and Stellmacher, G. 1985, A\&A 152, 175

\bibitem[Lorrain(1996)]{Lorrain1996}
Lorrain, P. and Koutchmy, S. 1996, Sol. Phys., 165, 115

\bibitem[November(1996)]{November1996}
November, L. and Koutchmy, S. 1996, ApJ, 466, 512

\bibitem[Tavabi(2015)]{Tavabi2015}
Tavabi, E., Koutchmy, S. and Ajabshirizadeh, A
{\sl Limb events brightenings and fast ejection using IRIS Mission Observation}
Accepted September $1^{st}$. 2015. in Sol. Phys. DOI: 10.1007/s11207-015-0771-3

\bibitem[Vernazza(1981)]{Vernazza1981}
Vernazza J. E., Avrett E. H. and Loeser, R. 1981, ApJ Suppl. Series, 45, 635-725

\end{thebibliography}

\end{document}